\documentstyle[preprint,aps,eqsecnum,epsf,prd,floats]{revtex}
\begin{document}
\preprint{\vbox{
\hbox{UTPT 96--14}
\hbox{UCSD/PTH 96--24}
\hbox{hep-ph/9610534}
}}
\title{Bound States and Power Counting \\ in Effective Field Theories}
\author{Michael Luke}
\address{Department of Physics, University of Toronto, \\
60 St. George St., Toronto, Ontario, Canada M5S1A7}
\author{Aneesh V.~Manohar}
\address{Department of Physics, University of California at San Diego,\\
9500 Gilman Drive, La Jolla, CA 92093-0319}
\date{October 1996}
\maketitle
\begin{abstract}
The problem of bound states in effective field theories is studied. A rescaled
version of nonrelativistic effective field  theory is formulated which makes
the velocity power counting of operators manifest.  Results obtained using the
rescaled theory are compared with known results from NRQCD. The same ideas are
then applied to study Yukawa bound states in $1+1$ and $3+1$  dimensions, and
to analyze when the Yukawa potential can be replaced by a $\delta$-function
potential. The implications of these results for the study of nucleon-nucleon
scattering in chiral perturbation theory is discussed.
\end{abstract}
\tighten

\section{Introduction}\label{sec:intro}

Effective field theories are an extremely useful tool for studying the dynamics
of particles at low energies. An effective Lagrangian typically has an
expansion in inverse powers of some mass scale $M$, and describes dynamics at
momentum scales which are much smaller than $M$. For example, heavy quark
effective theory (HQET) \cite{georgi,eh,iw} describes the dynamics of hadrons
containing a single heavy quark of mass $m_Q$ at momentum transfers much
smaller than $m_Q$. The HQET Lagrangian has an expansion in inverse powers of
$m_Q$, and is used to compute hadronic properties as an expansion in
${\Lambda_{\rm QCD}}/m_Q$, where ${\Lambda_{\rm QCD}} \sim 300$~MeV is a
typical strong interaction scale. The scale at which the HQET Lagrangian ceases
to be useful is the mass $m_Q$ of the heavy quark. HQET can be used to study
the interaction of a single heavy quark with light degrees of freedom, provided
the momentum transfer is small compared with $m_Q$.

Systems containing a heavy quark and antiquark (or two heavy quarks) can not be
described by HQET. As $m_Q \rightarrow \infty$, the quark and antiquark form a
Coulomb bound state of size $1/m_Q \alpha_s$, and typical momentum  transfer $p
\sim \alpha_s m_Q$. Perturbation theory is infrared divergent, with terms of
the form $\left( \alpha/v \right)^n$, where $v$ is the relative velocity of the
$Q$ and $\bar Q$ in the center of mass frame. These infrared singular terms
cause a breakdown of perturbation theory, and must be resummed.  Resummation of
the most  singular $\left( \alpha / v \right )^n$ terms is equivalent to
solving the  Schrodinger equation in a Coulomb potential, and the resummed $Q
\bar Q$  scattering  amplitude contains the corresponding bound state poles.
One can construct a different effective field theory, NRQCD (non-relativistic
QCD), that is appropriate for the study of $Q \bar Q$ bound states in QCD (or
its QED analog, NRQED for the study of positronium) \cite{cl,bbl}.  The terms
in the NRQCD Lagrangian are of the same form as those for HQET, but NRQCD has a
different power counting scheme than HQET. In HQET, the power counting of
operators is manifest. An operator with coefficient $1/m_Q^r$ has a matrix
element of order
$\left(\Lambda_{\rm QCD}/m_Q\right)^r$.  Thus, the quark kinetic energy
operator $Q^\dagger\left( {\bf D}^2/2m_Q\right) Q$  is of order $\Lambda_{\rm
QCD}/m_Q$, and is subleading. The power counting
is more subtle in NRQCD --- $Q^\dagger \left( {\bf D}^2/2m_Q \right) Q$ is
treated as a leading order operator, and higher dimension operators are
suppressed not by powers of $\Lambda_{\rm QCD}/m_Q$ but by powers of the
relative three-velocity of the heavy quarks.   We will refer to theories such
as HQET, in which $Q^\dagger\left( iD^0\right) Q$ is of leading order but
$Q^\dagger\left( {\bf D}^2/2 m_Q\right) Q$ is small, as  static theories, and
will refer to theories such as NRQCD, in which $Q^\dagger \left(i D^0\right) Q$
and $Q^\dagger\left(  {\bf D}^2/2 m_Q \right) Q$ are both of leading order, as
non-relativistic theories.

In this paper, we study the problem of non-relativistic bound states in
effective field theories. In Sec.~\ref{sec:nrqcd}, we
define a rescaled version of NRQCD (RNRQCD) in which
the power counting of operators is manifest. Most of the results of RNRQCD
follow trivially from well-known results in NRQCD, but there are a few
advantages, which are discussed in Sec.~\ref{sec:nrqcd}. In
Sec.~\ref{sec:yukawa} we analyse non-relativistic Yukawa bound states due to
the exchange of a massive scalar, in both $1+1$ and $3+1$ dimensions, and
compare the results to those in an effective theory in which the scalar is
integrated out and replaced by a four-Fermi interaction. Naively, the momentum
expansion of the four-Fermi effective theory has a convergence radius equal to
the mass of the scalar, but this is not necessarily the case when there are
weakly bound states. We study whether the effective theory can ``produce
its own bound state,''  that is, whether one can obtain a composite weakly
bound
state in an effective theory where the higher dimensional operators have small
coefficients. We show that in $1+1$ dimensions, the effective four-Fermi theory
has a radius of convergence of order the scalar mass, and produces a weakly
bound state with perturbative coefficients for higher dimension operators when
the parameters of the Yukawa theory are such that there is one weakly bound
state in the spectrum. If the Yukawa theory has two or more bound states of
which one is weakly bound, the radius of convergence of the four-Fermi
effective theory vanishes as the weakly bound state approaches threshold, and
the higher dimension operators have large coefficients. In $3+1$ dimensions,
the four-Fermi effective theory has a finite radius of convergence and higher
dimension operators with small coefficients only when there is no bound state
near threshold in the Yukawa theory. We discuss the implications of our
analysis for recent attempts to describe non-perturbative aspects of
nucleon-nucleon scattering by solving the Schrodinger equation for a chiral
Lagrangian in which the nucleon-nucleon interaction arises from both pions
exchange and contact terms~\cite{weinberg,vka,vkb,ksw}. We present our
conclusions in
Sec.~\ref{sec:conc}.

\section{Power Counting in Non-Relativistic QCD}\label{sec:nrqcd}

\begin{figure}
\epsfxsize=8cm
\hfil\epsfbox{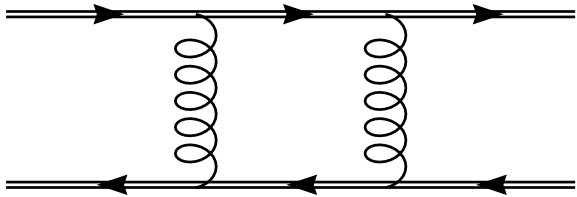}\hfill
\caption{The box graph in the $Q \bar Q$ sector. The double lines represent
nearly on-shell non-relativistic quarks and antiquarks, and the curly lines
represent gluons. \label{fig:qcdbox}}
\end{figure}

The HQET Lagrangian at leading order in the $1/m_Q$ expansion is
\begin{equation}\label{1.2}
{{\cal L}} = Q^\dagger \left( i D^0 \right) Q,
\end{equation}
where $Q$ is the annihilation field for a non-relativistic quark. It is well
known that the Lagrangian Eq.~(\ref{1.2}) cannot be used for systems
containing two heavy quarks. The basic problem arises from the Feynman graph in
Fig.~\ref{fig:qcdbox}, when both the intermediate fermions are simultaneously
almost on-shell. The box graph evaluated in QCD (with propagators
$1/\left({\rlap{$p$}/}+ m \right)$) has terms which are of order $m_Q/{\left|
{\bf k}\right|} = 1/\left|{ {\bf v}}\right|$, where ${\bf k}$ and ${\bf v}$ are
the  three-momentum and velocity of the external quarks, respectively. The box
diagrams in HQET can not reproduce this behavior since the Feynman rules are
independent of $m_Q$. As a result, QCD can not be matched on to HQET in the
$Q\bar Q$ sector. The box graph in HQET has a loop integral of the form
\begin{equation}\label{1.2.1}
\int {d^{4} {k} \over \left(2 \pi \right)^{4}} \ {1\over k^0 + i \epsilon}\ { 1
\over -k^0 + i \epsilon}.
\end{equation}
The $k^0$ integral has a pinch singularity, and is divergent.  This singularity
is a signal that higher dimension operators in HQET are important in the $Q
\bar Q$ sector. One constructs a different effective field theory, NRQCD,  with
the leading order quark Lagrangian
\begin{equation}\label{1.3}
{{\cal L}} = Q^\dagger\left( i D^0 \right) Q + Q^\dagger  \left( {{\bf D}^2
\over 2 m_Q} \right) Q + \ldots \ ,
\end{equation}
in which ${\bf D}^2/2m_Q$ is considered to be of the same order as $D^0$. The
box graph in NRQCD has a loop integral of the form
\begin{equation}\label{1.2.2.}
\int {d^{4} {k} \over \left(2 \pi \right)^{4}}\ {1\over k^0 - \left|{\bf
k}\right|^2/2m_Q + i \epsilon}\ { 1 \over -k^0 -\left|{\bf k}\right|^2/2 m_Q +
i \epsilon} \ \ldots\ ,
\end{equation}
instead of Eq.~(\ref{1.2.1}). The $k^0$ integral is now finite, and gives
\begin{equation}\label{1.2.3}
-i \int {d^{3} {k} \over \left(2 \pi \right)^{3}}\ {m_Q \over \left|{\bf
k}\right|^2}\ \ldots\ ,
\end{equation} which, when the gluon propagators are included,
reproduces the $m_Q/\left| {\bf k }\right|$  enhancement
of the box graph in QCD.

HQET has a simple power counting scheme in which $D^\mu$ is of order
${\Lambda_{\rm QCD}}$. The importance of various operators is then manifest
from the Lagrangian. An operator in the Lagrangian of dimension $4+r$ has a
coefficient of order $1/m_Q^r$, and is of relative order $\left({\Lambda_{\rm
QCD}}/m_Q\right)^r$. It is also trivial to count powers of $1/m_Q$ in loop
graphs. The quark and gluon propagators are independent of $m_Q$, so any
Feynman graph  with $V$ vertices of order $1/m_Q^{r_1} \ldots 1/m_Q^{r_V}$ has
an overall factor of $1/m_Q^r$, where $r=\sum_{i=1}^V r_i$. NRQCD has a more
complicated power counting scheme, which is discussed in detail in
Ref.~\cite{bbl}. In NRQCD, both $D^0$ and ${\bf D}^2/2m_Q$ are of the same
order, so the quark propagator is $i/\left(k^0 - {\bf k^2}/2m_Q+i
\epsilon\right)$. The quark propagator depends on $m_Q$, so that one can get
factors of $m_Q$ from loop graphs, and the power counting rules for loop graphs
in NRQCD is not as straightforward as in HQET.   Furthermore, there are several
relevant scales in NRQCD:  $m_Q$, the three-moment of the heavy quarks $m_Q v$,
the kinetic energy of the heavy quarks $m_Q v^2/2$, and $\Lambda_{\rm QCD}$.
The matrix elements of higher dimension operators in NRQCD are suppressed by
powers of $v$, but the $v$ counting is not manifest  in the Lagrangian.

It is advantageous to have an effective field theory with manifest power
counting rules. One can achieve this for NRQCD by rescaling the fields and
coordinates of the usual NRQCD Lagrangian.  In a nonrelativistic system, $E$
and $p$ are of order $m_Q v^2$ and $m_Q v$, respectively.  Therefore, it is
useful to rescale the coordinates so that these are the natural sizes of the
energy and momentum. Define new coordinates ${\bf X}$ and $T$, and new fields
$\Psi$, ${{\cal A}}^0$ and ${\bf{{\cal A}}}$ by
\begin{equation}
{\bf x} = \lambda_x {\bf X},\ \ t=\lambda_t T,\ Q = \lambda_Q \Psi,
\ A^0 = \lambda_{A^0} {{\cal A}}^0,\ {\bf A} = \lambda_{{\bf  A}}
{\bf {{\cal A}}}, \label{1.4}
\end{equation}
where $\lambda_x=1/m_Q v$. Requiring $\partial^0$ and ${\nabla}^2/2m_Q$ to be
both of the same order  determines $\lambda_t=m_Q \lambda_x^2=1/m_Q v^2$.  This
gives the relation between the rescaled energy and momentum $K^0$ and ${\bf K}$
and the original variables $k^0$ and ${\bf k}$,
\begin{eqnarray}
K^0&=&k^0/m_Q v^2,\nonumber \\ {\bf K}&=&{\bf k}/m_Q v.
\end{eqnarray}
In a nonrelativistic system the rescaled energy and momentum are both of order
unity.

Upon rescaling, the Lagrangian density picks up an overall factor of
$\lambda_x^3\lambda_t$ from the change of integration variables $d^3x\, dt
\rightarrow \lambda_x^3\lambda_t\, d^3X\, dT$, so the $\Psi^\dagger\left(i
D^0\right) \Psi$ term is canonically normalized if $\lambda_Q =
\lambda_x^{-3/2}$.  The gauge field quadratic terms get rescaled to
\begin{eqnarray}
\left({\bf \nabla \times A} \right)^2 &\rightarrow& m_Q \lambda_x^3 \lambda_A^2
\left({\bf \nabla} \times {\bf{{\cal A}}}  \right)^2 , \nonumber \\
\noalign{\medskip}
\left({\bf \nabla} A^0 \right)^2 &\rightarrow& m_Q \lambda_x^3 \lambda_{A^0}^2
\left({\bf \nabla {{\cal A}}^0 } \right)^2 ,  \nonumber\\
\noalign{\medskip}
\left({\partial {\bf A} \over \partial t} \right)^2 &\rightarrow& {\lambda_x
\lambda_{A}^2\over m_Q} \left({\partial {\bf {{\cal A}}}\over  \partial t}
\right)^2, \label{1.5} \\
\noalign{\medskip}
{\partial {\bf A}\over \partial t}\cdot {\bf \nabla} A^0  &\rightarrow&
\lambda_x^2 \lambda_{A^0} \lambda_{A} {\partial {\bf A}\over \partial t}\cdot
{\bf \nabla} A^0. \nonumber
\end{eqnarray}
In the infrared, $\lambda_x \rightarrow \infty$, and the dominant gauge kinetic
terms are $\left({\bf \nabla \times A} \right)^2$ and  $\left({\bf \nabla} A^0
\right)^2$. These terms are properly normalized if
\begin{equation}\label{1.6}
\lambda_A = \lambda_{A^0} = \left( m_Q \lambda_x^3 \right)^{-1/2}.
\end{equation}
With this rescaling the NRQCD Lagrangian becomes
\begin{eqnarray}
{{\cal L}}^R &=&\Psi^\dagger \left(i\partial_0 - {g\over \sqrt{ v }} {{\cal
A}}_0^a T^a \right) \Psi \nonumber- {1\over2} \Psi^\dagger \left(i\nabla - g
\sqrt{v}
{\bf{{\cal A}}}^a T^a \right)^2 \Psi  - {1\over 4} \left( \partial_i {{\cal
A}}_j^a -
\partial_j {{\cal A}}_i^a -  g \sqrt{v} f_{abc} {{\cal A}}_i^b {{\cal A}}_j^c
\right)^2 \nonumber \\
&&+ {1\over 2} \left(\partial_i {{\cal A}}_0^a - v  \partial_0 {{\cal A}}_i^a -
g \sqrt{v} f_{abc} {{\cal A}}_i^b {{\cal A}}_0^c \right)^2 \nonumber \\
&=&  \Psi^\dagger \left(i\partial_0 + {\nabla^2\over 2}-{g\over \sqrt{ v }}
{{\cal A}}_0^a T^a \right)\Psi  - {1\over 4} \left( \partial_i {{\cal A}}_j^a -
\partial_j {{\cal A}}_i^a\right)^2+  {1\over 2} \left(\partial_i {{\cal
A}}_0^a\right)^2+{{\cal O}}(v, g\sqrt{v})  \label{rescaled}
\end{eqnarray}
which will be referred to as the RNRQCD (rescaled NRQCD) Lagrangian.
(The effective Lagrangian also contains the corresponding terms for the heavy
antiquark
field.)

It is clear from the form of the RNRQCD Lagrangian that the effective coupling
constant in the $\Psi\Psi$ sector for Coulomb gluons (i.e ${{\cal A}}^0$) is
$\alpha/v$, not $\alpha$. At low velocities, the Coulomb gluon interaction must
be summed to all orders. Transverse gluons have a coupling constant of order $g
\sqrt v$, and decouple as $v \rightarrow 0$. Loop integrals with Coulomb gluons
are independent of $g$, $v$ and $m_Q$, so the $v$ dependence of a graph may be
easily read off from the vertex  factors.  There are no hidden enhancements
factors from loop graphs.\footnote{$v$ counting for transverse gluons is more
complicated, and is discussed in Sec.~\ref{sub:tg}.} This is in contrast with
the usual formulation of NRQCD, where $1/v$ enhancements arise due to factors
of $m_Q/|{\bf k}|$ in the loop graphs. For non-relativistic bound states, $v$
is of order $\alpha$, and the Coulomb interaction in  Eq.~(\ref{rescaled})
becomes strongly coupled and must be summed to all orders.   Other
interactions, such as those due to transverse gluons, are suppressed by powers
of $v$ and may be treated using perturbation theory.   These well-known results
follow simply from the scaling of the various terms in  the rescaled Lagrangian
Eq.~(\ref{rescaled}).  The derivation using the original NRQCD Lagrangian is
more involved \cite{bbl,labelle}.

The effective Lagrangian Eq.\ (\ref{rescaled}) contains additional higher
dimension operators which are also suppressed by factors of $v$.  Terms in the
effective Lagrangian are relevant, irrelevant or marginal, depending on whether
the power of $v$ in the coefficient is negative, positive or
zero.\footnote{This is in the renormalization group sense. Irrelevant operators
inserted in loop graphs which are sufficiently divergent can produce effects
that do not vanish as $v \rightarrow 0$.} Note that in RNRQCD  the fields and
derivatives are all dimensionless, so terms with additional fields or
derivatives are not suppressed unless they appear with additional powers of
$v$. For example, the operator
\begin{equation}
{1\over m_Q^3} \psi^\dagger \nabla^4\psi
\end{equation}
in NRQCD becomes, in the rescaled theory,
\begin{equation}
{1\over m_Q^3}\,{m_Q\lambda_x^5\lambda_Q^2\over \lambda_x^4}\ \Psi ^ \dagger
\nabla^4\Psi= v^2\, \bar\Psi\nabla^4\Psi,
\end{equation}
and is of order $v^2$, which agrees with the power counting in
Ref.~\cite{bbl}.  The relation between our power counting and that of
Ref.~\cite{bbl} is slightly more subtle for operators containing $\vec E$ and
$\vec B$ fields.  For example, the chromomagnetic moment operator
\begin{equation}
{g\over m_Q}\,\psi^\dagger\,\sigma^{\mu\nu}G_{\mu\nu}\psi
\end{equation}
becomes, in the rescaled theory
\begin{equation}\label{chromo}
{g\over m_Q}\, m_Q\lambda^4\lambda_Q^2\lambda_A\ \Psi^\dagger\sigma^{\mu\nu}
{\cal G}_{\mu\nu}\Psi=g\sqrt{v}\, \Psi^\dagger\sigma^{\mu\nu} {\cal
G}_{\mu\nu}\Psi
\end{equation}
whereas in the power counting of Ref.~\cite{bbl} this operator is of order
$v^2$. However, the NRQCD power counting refers to the size of the matrix
element of the operator in a quark-antiquark state.  The chromomagnetic gluon
must
therefore be attached to one of the external
quark lines, which costs an additional power of $g\sqrt{v}$; hence the matrix
element of the operator Eq.~(\ref{chromo})  is of order $g^2 v\sim v^2$ in a
quarkonium state, as expected.

\subsection{Tranverse Gluons}\label{sub:tg}

\begin{figure}
\epsfxsize=7cm
\hfil\epsfbox{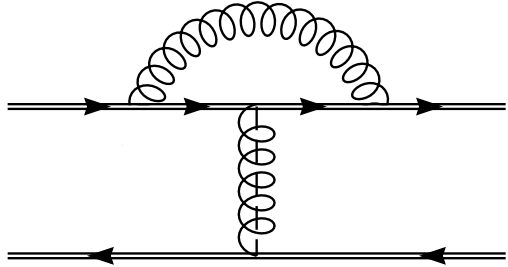}\hfill
\caption{Radiative correction to Coulomb scattering. The gluon with a dashed
line is a Coulomb gluon.\label{fig:qcdloop}}
\end{figure}

The quark propagator in RNRQCD is
\begin{equation}\label{1.7}
{i\over K^0 - \left|{\bf K}\right|^2/2 + i \epsilon},
\end{equation}
and the propagator for $A^0$ is
\begin{equation}\label{1.8}
{i \over \left|{\bf K}\right|^2},
\end{equation}
which are both independent of $v$. The transverse gluon propagator is
\begin{equation}\label{1.9}
-i\left(\delta_{ij} - {K_i K_j \over {\left|{\bf K}\right|^2}}\right) {1 \over
\left|{\bf K}\right|^2-v^2\left(K^0\right)^2}.
\end{equation}
which depends on $v$, so there are potential $1/v$ enhancements from graphs
with internal transverse gluons. A generic loop integral can be evaluated by
first doing the $K^0$ integral  using residues. The residue of the transverse
gluon propagator at the $K^0$ pole $\left|{\bf K} \right|  /v$ is $-1/v
\left|{\bf K} \right|$, which is enhanced by $1/v$. The  transverse gluon
propagator contribution from other poles (such as fermion poles) is typically
of order unity, since these poles are at values of $K^0$ of order unity. Thus
transverse gluon loops can have $1/v$ enhancements from
regions in the momentum integral where the transverse gluon is on-shell.
Consider, for example, the graph in Fig.~\ref{fig:qcdloop}, where the gluon in
the loop is a transverse gluon. The diagram has a $1/v$ enhancement from the
region in loop integral where the fermion and virtual gluon are on-shell, i.e.
from physical tranverse gluon radiation.

The transverse gluon coupling constant is $g \sqrt v$, which is a factor of $v$
smaller than that for Coulomb gluons.  Thus transverse gluon loops obey naive
$v$ counting, and are $v^2$ suppressed, unless the cut part of the graph
contributes to transverse gluon radiation. In the latter case, the transverse
gluon graph has a $1/v$ enhancement over the naive $v$-counting, and is only
suppressed by one power of $v$.

It is not possible to choose a rescaling scheme which has a manifest
$v$-counting scheme for real and virtual transverse gluons. The reason is that
quarks behave like non-relativistic particles when transverse gluons are
exchanged between them, but like static particles when one of them radiates an
on-shell transverse gluon.

\subsection{Coulomb Scattering}\label{sub:coulomb}

\begin{figure}
\epsfxsize=14cm
\hfil\epsfbox{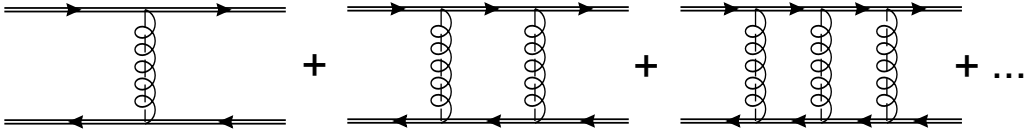}\hfill
\caption{The leading order contribution to $Q \bar Q$ scattering. The gluons
with dashed lines represent Coulomb gluons. \label{fig:qcdladder}}
\end{figure}

\begin{figure}
\epsfxsize=7cm
\hfil\epsfbox{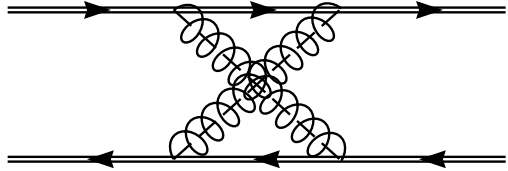}\hfill
\caption{The crossed box graph.\label{fig:qcdcross}}
\end{figure}

At leading order in $v$,  the only diagrams which contribute to $\bar Q Q$
scattering in the effective theory  are the Coulomb ladder graphs of
Fig.~\ref{fig:qcdladder},  where the ${{\cal A}}^0$ propagators are denoted by
gluons with dashed lines. Crossed ladder graphs such as Fig.~\ref{fig:qcdcross}
vanish when both gluons are Coulomb gluons. The one-loop box graph has an
integral  of the form
\begin{eqnarray}
&&\int {d^{} {K^0} \over \left(2 \pi \right)^{}} {d^{3} {{\bf K}} \over \left(2
\pi \right)^{3}} \ {1\over {\left|{\bf K}\right|}^2} {1\over  \left|{\bf
K}+{\bf P_1}^\prime - {\bf P_1} \right|^2} \ {1\over \left(P_1 - K \right)^0 -
\left|{\bf P_1} - {\bf K} \right|^2/2 + i \epsilon} \times \label{3.12} \\
&&\qquad{1\over \left(P_2+ K \right)^0 - \left|{\bf P_2} + {\bf K} \right|^2/2
+ i \epsilon}. \nonumber
\end{eqnarray}
The $K^0$ integral can be done by contour integration to give
\begin{eqnarray}
&&\int {d^{} {K^0} \over \left(2 \pi \right)^{}} {1\over \left(P_1 - K
\right)^0 - \left|{\bf P_1} - {\bf K} \right|^2/2 + i \epsilon} \ {1\over
\left(P_2 + K\right)^0 - \left|{\bf P_2} + {\bf K} \right|^2/2 + i \epsilon}
\label{3.13} \\
&&= -{i\over \left( P_1 +P_2 \right)^0 - \left|{\bf P_1} - {\bf K} \right|^2/2
- \left|{\bf P_2} + {\bf K} \right|^2/2} \nonumber
\end{eqnarray}
which is the Schrodinger Green function of non-relativistic quantum mechanics.
The ladder graphs of Fig.~\ref{fig:qcdladder} can  be evaluated by first doing
the $K^0$ loop integrals.  It is easy to see that each loop gives the
Schrodinger Green function for the intermediate two-fermion state. The sum of
all the ladder graphs gives the non-relativistic Schrodinger equation in
momentum space for a Coulomb potential. This result is, of course, well-known.
What is different here is that in the rescaled theory, the sum of the leading
order graphs is identical to the Schrodinger equation. While this is also the
case for NRQCD in Coulomb gauge, the rescaling allows similar results to be
obtained in non-gauge theories, such as Yukawa theory (which we will discuss at
length in the next section), where there is no gauge freedom in choosing the
propagator.

\subsection{Heavy-Light Systems}\label{sub:heavylight}

One can see explicitly why ${\bf D}^2/2 m_Q$ is a relevant operator for
heavy-heavy bound states, but not for heavy-light bound states. Consider a
non-relativistic bound state of two particles with different masses $m_H$ and
$m_L$, with $m_H \gg m_L$. Applying the rescaling Eq.~(\ref{1.6}) with $m_Q
\rightarrow m_L$, one finds that the fermion kinetic terms are
\begin{equation}
\psi_L^\dagger\, i\partial^0\, \psi_L + {1\over 2}\,
\psi_L^\dagger  {\bf \nabla}^2 \psi_L
\end{equation}
for the light fermion, and
\begin{equation}
\psi_H^\dagger\, i\partial^0 \psi_H + {m_L \over 2 m_H}\,
\psi_H^\dagger {\bf \nabla}^2 \psi_H
\end{equation}
for the heavy fermion. The $\psi^\dagger  {\bf \nabla}^2 \psi$ operator is
comparable to $\psi^\dagger \partial^0 \psi$ for the light particle, but is
smaller by $m_L/m_H$ for the heavier particle. The heavier particle can be
treated as a static source (as in HQET), but the lighter particle must be
treated as non-relativistic (as in NRQCD).

\subsection{Scaling Dimensions}\label{sub:modscal}

The velocity scaling rules have a renormalization group interpretation. In the
non-relativistic limit, the scaling dimensions of space and time should be
chosen to be
\begin{equation}\label{1.50}
\left[{\bf x}\right]=-1,\ \left[t\right]=-2,
\end{equation}
instead of the usual choice in relativistic theories
\begin{equation}
\left[{\bf x}\right]=-1,\ \left[t\right]=-1.
\end{equation}
With the choice Eq.~(\ref{1.50}), one finds that
\begin{eqnarray}
&&\left[{{\cal L}}\right]=5,\\
&&\left[\psi\right]=3/2,\ \left[{{\cal A}}^0\right]=3/2,\
\left[{\bf{{\cal A}}}\right]=3/2, \label{1.10b}\\
&&\left[k^0\right]=2,\ \left[{\bf k}\right] = 1 .\nonumber
\end{eqnarray}
Operators are relevant, irrelevant, or marginal depending on whether their
dimensions are less than, greater than, or equal to $\left[{{\cal
L}}\right]=5$. For example, the Coulomb interaction term $\Psi^\dagger A^0
\Psi$ has dimension $9/2=5-1/2$, so the Coulomb interaction is relevant, and of
order $v^{-1/2}$. The transverse gluon interaction term is $\Psi^\dagger {\bf p
\cdot A} \Psi$ and has dimension $11/2=5+1/2$, so the term is irrelevant, and
of order  $v^{1/2}$. This agrees with the powers of $v$ in the Lagrangian
Eq.~(\ref{rescaled}).

\section{Power Counting in Chiral Perturbation Theory}\label{sec:yukawa}

There has been much recent interest in applying the techniques of
non-relativistic  effective theories to nucleon-nucleon
scattering \cite{weinberg,vka,vkb,ksw}.  The goal is to describe low-energy
nucleon-nucleon scattering using a chiral Lagrangian where massive excitations
(such as $\rho$'s and $\omega$'s) have
been integrated out and replaced by an effective four point interaction.
Nonperturbative effects, such as the
large scattering length in the $^1S_0$ channel or the deuteron in the $^3S_1$
channel, could be described by solving the Schrodinger equation in the
effective theory.

The rescaling of the previous section can be extended to chiral perturbation
theory for heavy nucleons \cite{jm,weinberg,ksw}.  The leading terms in the
nucleon-pion chiral Lagrangian have the form
\begin{eqnarray}\label{4.1}
&& {{\cal L}} =  \psi^\dagger\, \left(i \partial^0 + { {\bf
\nabla}^2 \over 2 M } \right)\, \psi - {g\over f_\pi} \,
{\bf \nabla}\phi\cdot\psi^\dagger
{\bf \sigma} \psi  \\
&&+ {1\over 2} \left(\partial^0 \phi \right)^2 -  {1\over 2} \left({\bf \nabla}
\phi \right)^2-{c_1\over f_\pi^2}\left(\psi^\dagger\psi\right)^2 -{c_2\over
f_\pi^2}\left(\psi^\dagger{\bf \sigma}\psi\right)^2\nonumber
\end{eqnarray}
describing the interaction of a pseudoscalar Goldstone boson $\phi$ with decay
constant $f_\pi$  and coupling $g$, where we have suppressed flavour indices.
The leading effects of massive excitations are contained in the dimension six
operators $(\psi^\dagger\psi)^2$ and $(\psi^\dagger{\bf\sigma}\psi)^2$ with
dimensionless coefficients $c_1$ and  $c_2$.  In $3+1$ dimensions the
appropriate rescaling is
\begin{equation}\label{again3.24}
{\bf x} = \lambda_x {\bf X},\ t = M \lambda_x^2 T, \ \psi = \lambda_x^{-3/2}
\Psi,\ \phi =  \left( M \lambda_x^3 \right)^{-1/2}  \Phi
\end{equation}
giving the rescaled Lagrangian
\begin{eqnarray}\label{4.2}
&& {{\cal L}} =  \Psi^\dagger \left(i \partial^0 +  { {\bf
\nabla}^2 \over 2}\right) \Psi - {g M \sqrt v\over f _\pi}\, {\bf
\nabla}\Phi\cdot
\Psi^\dagger{\bf \sigma} \Psi  \\
&&+ {v^2\over 2 }  \left(\partial^0 \Phi \right)^2 -  {1\over 2} \left({\bf
\nabla} \Phi \right)^2 -{c_1 M^2 v\over
f_\pi^2}\left({\Psi^\dagger\Psi}\right)^2
-{c_2 M^2 v\over f_\pi^2}\left({\Psi^\dagger{\bf\sigma}\Psi}\right)^2.\nonumber
\end{eqnarray}
Both the derivative interaction and the dimension six operators are  irrelevant
in the infrared. Each pion exchange in a Feynman diagram contributes a factor
of $g^2 M^2 v/f_\pi^2$, which has the same dependence of $M$ and $v$ as an
insertion of one of the dimension six four-nucleon operators. This reproduces
the power counting of Ref.~\cite{weinberg}: a graph with $n$ insertions of
dimension six operators and $m$ ladder pion exchanges scales as
\begin{equation}\label{mpower}
\left(M^2 v\over f_\pi^2\right)^{n+m}=\left(M Q\over f_\pi^2\right)^{n+m}
\end{equation}
where $Q=M v$ is the three momentum of the scattering nucleons (in the  centre
of mass frame). Furthermore, with this rescaling the power counting of Ref.
\cite{ksw}, in which poles in $k^0$ in the pion propagator were treated as
higher order in the nonrelativistic expansion, is manifest.  Since pion
emission
is kinematically forbidden as $v\rightarrow 0$, the temporal piece of the pion
kinetic
term may be treated as a higher-order insertion.

Since the pion-nucleon interaction terms are irrelevant in the infrared, unlike
the interactions in QED and QCD, perturbation theory for nucleon-nucleon
scattering does not break down at threshold for weak coupling. Multi-loop
bubble graphs are not enhanced by powers of $1/v$, but instead suppressed by
powers of $v$. It was argued in Refs.~\cite{weinberg,vka,vkb,ksw}
that the large scattering length in the $^1 S_0$ channel and the deuteron in
the $^3 S_1$ channel in nucleon-nucleon scattering are signs of the breakdown
of perturbation theory, due to the fact that for large
$M$ the factor in Eq.~(\ref{mpower}) is not small. It was proposed by these
authors that the appropriate description of nucleon-nucleon scattering could be
obtained by summing all terms of order $(QM)^n$; corrections to this would be
suppressed by powers of $Q/M$, and could be calculated systematically.

However, as noted in Ref.~\cite{ksw}, the validity of this approach rests on
the  assumption that coefficients of higher dimension four-nucleon operators in
the chiral Lagrangian with $n$ spatial derivatives are smaller than $M^n$.
Otherwise, the effective theory description would break down, since the
contributions from higher dimension operators with unknown coefficients would
be as large as the  graphs which are being summed in the effective theory. In
\cite{ksw} it was argued that this assumption was consistent in the $^1 S_0$
(but not in the $^3 S_1$) channel, since no divergences were found in the
Feynman diagrams contributing to $^1S_0$ channel scattering which would require
counterterms scaling like $(QM)^n$.

In this section we will investigate the validity of this power counting in a
simplified model. We will neglect the Goldstone bosons, and simply consider a
theory of a non-relativistic fermion of mass $M$ coupled to a scalar of mass
$m\ll M$,
\begin{equation}\label{3.1}
{{\cal L}}_Y =  \psi^\dagger\left( i \partial^0 + { { \nabla}^2 \over 2M
}\right)  \psi - g\, \phi \psi^\dagger \psi + {1\over 2} \left(\partial^0 \phi
\right)^2 -  {1\over 2} \left({\bf \nabla} \phi \right)^2 - {1\over 2} m^2
\phi^2.\nonumber
\end{equation}
The scalar plays the role of the $\rho$, $\omega$ and other excitations.  At
low momenta, $p\ll m$, the scalar may be integrated out, resulting in an
effective four-fermi theory
\begin{equation}\label{ffermi}
{{\cal L}}_\delta =  \psi^\dagger\left( i \partial^0 + { { \nabla}^2 \over 2M
}\right)  \psi + h\left(\psi^\dagger \psi\right)^2+\dots
\end{equation}
where the ellipses represent higher dimension operators, suppressed by powers
of $p/m$. The Yukawa potential arising from scalar exchange in Eq. (\ref{3.1})
is replaced by a series of delta functions and their derivatives in the
effective theory. We will refer to these two theories as ${{\rm NRY}}$ and
${{\rm NR}\delta}$ for non-relativistic Yukawa, and non-relativistic
$\delta$-function, respectively.

We now wish to ask whether ${{\rm NR}\delta}$ correctly describes the
$\psi\psi$ scattering amplitude of ${{\rm NRY}}$ at low momentum $p \le p_{\rm
max}$, in a parameter regime where the Born approximation fails.  By low
momentum,
we mean $p \le p_{\rm max} \ll m$, where $p_{\rm max}$ is held fixed as one
varies the parameters in the Yukawa theory. This is analogous to the
question of whether the dimension six operators in Eq.~(\ref{4.2}) correctly
describe nucleon-nucleon scattering at momenta much smaller than $m_\rho$, in a
regime where the Born approximation fails. We consider the problem in both
$1+1$ and $3+1$ dimensions.  In 1+1 dimensions, we will show that ${{\rm
NR}\delta}$ correctly describes the scattering for weak coupling; however, for
a coupling large enough that more than one bound state exists the higher
dimension operators in ${{\rm NR}\delta}$ become important, and the effective
theory breaks down below $p_{\rm max}$.
In this case, the full theory is required to correctly describe
the low-energy physics.  In 3+1 dimensions, there are no bound states for
sufficiently weak coupling. In
this case ${{\rm NR}\delta}$ correctly describes the scattering and no
resummation of ladder graphs is necessary.  However, when the coupling is
strong enough to form bound states, the situation is analogous to the 1+1
dimensional case with excited states. Once again the higher dimension
operators become important, and again the effective theory description
breaks down below $p_{\rm max}$. We find by explicit calculation that there
are,
in fact, higher
dimension operators in the low energy theory which scale like $(QM)^n$, and
spoil the power counting of Refs.~\cite{weinberg,vka,vkb,ksw}.  We will
comment on the relation between our results and the effective range expansion,
as well as the case where $p_{\rm max}$ is allowed to vary, in
Sec.~\ref{sub:ere}.

\subsection{$1+1$ Dimension}\label{sub:1d}

In $1+1$ dimensions, the mass dimensions of the fields and couplings in
Eq.~(\ref{3.1}) are
\begin{equation}\label{3.1b}
\left[ \psi \right ] =1/2,\ \left[ \phi \right] =0, \ \left[ M \right] = 1,\
\left[ m \right] = 1,
\ \left[ g \right] = 1.
\end{equation}
The Yukawa potential due to scalar exchange is
\begin{equation}\label{3.2}
V(x) = -g^2 \int {d^{} {q} \over \left(2 \pi \right)^{}} { e^ {i q x} \over q^2
+ m^2 } = -{g^2\over 2 m} e^{- m \left|x\right| }.
\end{equation}
An attractive potential in $1+1$ dimensions always has a bound state. The
Schrodinger equation for a Yukawa potential is not analytically  soluble.
However, when $g$ is small, the state is weakly bound and spread out over  a
large region. In this limit the Yukawa potential can be approximated by the
$\delta$-function,
\begin{equation}\label{3.3}
-{g^2\over 2 m} e^{- m \left|x\right| } \rightarrow -{g^2\over m^2} \delta(x)
\end{equation}
The $\delta$-function potential has a bound state with energy $E = -g^4 M/4
m^4$, and wavefunction
\begin{equation}\label{3.4}
\psi( x ) = \sqrt{ \kappa } e^{- \kappa \left| x \right| },\ \kappa
= {g^2 M \over m^2}.
\end{equation}
Corrections to the $\delta$-function result will be suppressed by powers of
$\kappa/m$, the ratio of the size of the state to the range of the potential.
For $\kappa/m\sim 1$, excited bound states appear, and the $\delta$-function
approximation completely breaks down.

In the effective field theory language, this indicates that
${{\rm NR}\delta}$ will only correctly describe bound states for weak coupling,
$\kappa/m\ll 1$, and will break down at larger values of the coupling.  Since
an excited state may be arbitrarily close to threshold, the effective theory
may therefore break down for $\psi\psi$ scattering at arbitrarily low energies,
which is not the usual behavior one expects from a low-energy effective theory,
where the importance of higher dimension operators is set by powers of $p/m$.
In this case, the effects of higher dimension operators must be suppressed
instead by powers of $\kappa/m$. Therefore, it is instructive to analyze this
problem by matching ${{\rm NRY}}$ onto ${{\rm NR}\delta}$  and seeing how the
scale $\kappa$ enters the problem.

\begin{figure}
\epsfxsize=14cm
\hfil\epsfbox{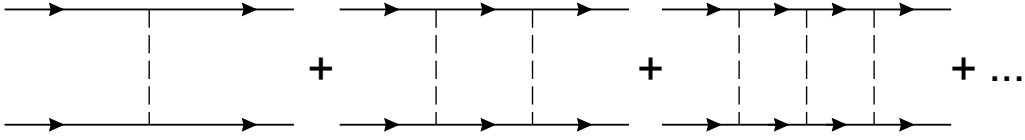}\hfill
\caption{Ladder graphs in the Yukawa theory.\label{fig:philadder}}
\end{figure}

To make the power counting manifest we use a rescaled Lagrangian as in
Sect.~\ref{sec:nrqcd}. In 1+1 dimensions, the appropriate rescaling is
\begin{equation}
x=\lambda_x X,\ t=M\lambda_x^2 T,\ \psi=\lambda_x^{-1/2}\Psi,\ \phi
=(M\lambda_x)^{-1/2}\Phi.
\end{equation}
where $\lambda_x=1/Mv$, giving
\begin{equation}\label{3.6}
{{\cal L}}_Y^R = \Psi^\dagger\left( i \partial^0+{ {\nabla}^2  \over 2 }\right)
\Psi   - {g\over M}v^{-3/2} \Phi  \Psi^\dagger \Psi   + {v^2\over 2}
\left(\partial^0 \Phi \right)^2 -  {1\over 2} \left({\bf \nabla} \Phi \right)^2
- {1\over 2} {m^2\over M^2}v^{-2} \Phi^2.
\end{equation}
The powers of $v$ can also be obtained using modified scaling rules, as in
Sec.~\ref{sub:modscal}. The modified scaling dimensions are
\begin{equation}
\left[ x \right ] =- 1,\ \left[ t \right ] = -2,\  \left[ {{\cal L}} \right ]
= 3, \ \left[ \psi \right ] =1/2,\ \left[ \phi \right] =1/2. \label{3.6.1}
\end{equation}
$\left[\psi^\dagger \psi \phi \right]-\left[ {{\cal L}} \right] =-3/2$ and
$\left[ \phi^2\right] - \left[ {{\cal L}} \right ] = -2$, so that these terms
scale as $v^{-3/2}$ and $v^{-2}$, respectively. Since the temporal piece
of the scalar kinetic term
in Eq.~(\ref{3.6}) is of order $v^2$, it may be neglected at the order at which
we are working. Thus ${{\rm RNRY}}$ (rescaled ${{\rm NRY}}$) describes a
fermion coupled to a  static scalar field, with coupling constant
$gv^{-3/2}/M$. The $\Phi$ interactions must be summed to all orders, since the
Yukawa coupling is large. Because the $\Phi$
propagator has no poles in $K^0$ (emission of scalars is kinematically
forbidden as $v\rightarrow 0$) only ladder graphs are non-vanishing in the
nonrelativistic theory.
For simplicity we consider the scattering of two off-shell fermions with energy
$E$ and zero momentum, since the scattering amplitude will have the same
singularities as for on-shell states.   We will also treat the $\psi$'s as
distinguishable,
since the additional graphs from Fermi statistics are not relevant to our
arguments.
The tree-level
$\Phi$ exchange graph in the ladder sum in Fig.~\ref{fig:philadder} gives
\begin{eqnarray}
I_0^Y
&=&i{g^2\over M^2 v^3}{1\over\mu^2}\nonumber \\
&=&i{g^2\over m^2 v} \label{phitree}
\end{eqnarray}
where $\mu^2=m^2/M^2 v^2$. The box graph gives
\begin{eqnarray} I_1^Y&=&-i{g^4\over M^4 v^6} \int {d^{} {K} \over \left(2 \pi
\right)^{}} {1\over  (K^2 + \mu^2)^2}{1 \over \left(P_1 + P_2 \right)^0 -
K^2} \label{philoopgraph}\nonumber\\
&=& i {g^4\over M^4 v^6}\ { 2\mu + \sqrt \varepsilon \over 4 \mu^3 \sqrt
\varepsilon \left(\mu + \sqrt \varepsilon \right)^2 }\nonumber\\
&=& i{ g^4  \over  2  m^4 v^2 \varepsilon^{1/2}}  -i{3 g^4 M\over 4 m^5 v}+
{\cal O}(v^0)
\end{eqnarray}
where $-\varepsilon =\left( P_1 + P_2 \right)^0$ is the total rescaled energy
of the incoming particles. The two-loop box graph gives
\begin{eqnarray} I_2^Y&=&i{g^6\over M^6 v^9} \int {d^{} {K} \over \left(2 \pi
\right)^{}}\int {d^{} {L} \over \left(2 \pi
\right)^{}} {1\over  (K^2 + \mu^2)}
{1\over  (L^2 + \mu^2)}{1\over  (\left(K-L\right)^2 + \mu^2)}\nonumber\\
&&\qquad \times
{1 \over \left(P_1 + P_2 \right)^0 -
K^2}{1 \over \left(P_1 + P_2 \right)^0 -
L^2}  \label{phi2loop}\nonumber\\
&=& i {g^6\over M^6 v^9}\ { 6 \mu^2 + 9 \mu \sqrt \varepsilon +
2 \varepsilon \over 12 \mu^4
\varepsilon \left(\mu + \sqrt \varepsilon \right)^2
\left(\mu + 2 \sqrt \varepsilon \right) \left(2 \mu + \sqrt \varepsilon
\right)}\nonumber\\
&=& i{ g^6 \over  4  m^6 v^3 \varepsilon} -
i{ 3 g^6 M \over  4  m^7 v^2 \sqrt \varepsilon} +
i{ 41 g^6 M^2 \over  24  m^8 v} +
{\cal O}(v^0).
\end{eqnarray}

The ${{\cal O}}(1/v)$ term from the box graph Eq.~(\ref{philoopgraph}) is down
by a factor of $\kappa/m$ relative to the ${{\cal O}}(1/v)$ term due to
tree-level exchange, and so may be neglected in the weak coupling limit,
$\kappa\ll m$.  In this limit, one may sum the most singular contributions of
the graphs in Fig.~\ref{fig:philadder} to obtain for the ladder sum
\begin{eqnarray}\label{yladdersum}
i{\cal A}&=&i{g^2 \over m^2 v} \left[ 1 + {g^2 \varepsilon^{-1/2} \over 2  m^2
v} +\left(
{g^2 \varepsilon^{-1/2}\over 2 m^2 v}\right)^2 + \ldots \right] \nonumber\\
\noalign{\medskip}
&=&i{g^2\over m^2 v}\ {1\over 1 - g^2\varepsilon^{-1/2} /\left(2  m^2
v\right)}, \label{3.15}
\end{eqnarray}
which has a pole at $\varepsilon  = g^4/\left( 4 m^4 v^2\right)$.  Rescaling
back to physical units gives $E = - g^4 M/(4 m^4)$, which is the correct bound
state energy for $\kappa/m\ll 1$. There is always at least one bound state
pole, even for weak coupling, because the box graph diverges as
$v \rightarrow 0$.

When $\kappa/m$ is not small, the ladder sum is no longer a geometric series,
since an $n$-loop ladder graph has singularities of the form
\begin{equation}\label{kscale}
\left(\kappa\over m\right)^n{1\over v},\ \left(\kappa\over
m\right)^{n-1}{1\over
v}{1\over v \sqrt \varepsilon},
\ \dots\ \left(\kappa\over m\right){1\over v}
{1\over \left(v \sqrt\varepsilon\right)^{n-1}},\
{1\over v}{1\over \left(v \sqrt\varepsilon\right)^{n}},
\end{equation}
(ignoring factors of $g/m$) as can easily be verified by a
dimensional estimate of the $n$-loop graph.  This behavior can also be seen
from
the explicit computations Eqs.~(\ref{phitree}--\ref{phi2loop}). Each term in
Eq.~(\ref{kscale}) is as large as a term in the sum Eq.~(\ref{yladdersum}), and
can not be neglected. The scattering amplitude will be given  by  the Green
function for a Yukawa potential (which is not known analytically) and will have
poles at the energies of all the bound states.

\begin{figure}
\epsfxsize=12cm
\hfil\epsfbox{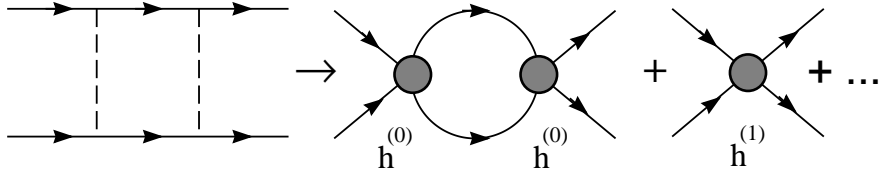}\hfill
\caption{One loop matching condition from ${{\rm NRY}}$ to ${{\rm NR}\delta}$.
\label{fig:loop}}
\end{figure}

\begin{figure}
\epsfxsize=14cm
\hfil\epsfbox{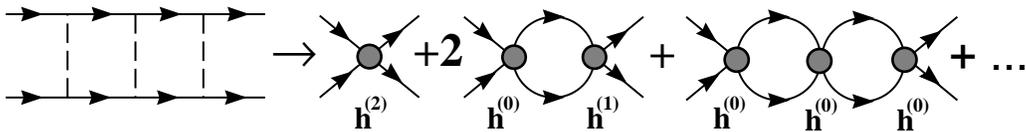}\hfill
\caption{Two loop matching condition from ${{\rm NRY}}$ to ${{\rm NR}\delta}$.
\label{fig:twoloop}}
\end{figure}

We now wish to integrate the $\Phi$ field out of the theory to obtain the
rescaled effective four-Fermi theory\footnote{The $v$ dependence of
$\left(\Psi^\dagger \Psi \right)^2$ can also be obtained using
Eq.~(\ref{3.6.1}), since $\left[ \left(\Psi^\dagger \Psi \right)^2 \right] -
\left[ {{\cal L}} \right ] =-1$.}
\begin{equation}\label{3.17}
{{\cal L}}_\delta^R = \Psi^\dagger\left( i \partial^0 + { { \nabla}^2 \over 2
}\right)  \Psi + {h\over v} \left(\Psi^\dagger \Psi \right)^2+ ....\ \ .
\end{equation}
The tree level matching condition arises from $\Phi$ exchange and gives
\begin{equation}\label{3.18}
h^{(0)} = {g^2 \over m^2}
\end{equation}
where we denote by $h^{(r)}$ the $r$-loop contribution to $h$. The one-loop
matching condition is shown in Fig.~\ref{fig:loop}.  The box graph has already
been evaluated in the full theory in  Eq.~(\ref{philoopgraph}).  In the
effective theory, the box graph is
\begin{eqnarray}\label{3.19}
I_1^\delta&=&\left({h^{(0)}\over v}\right)^2 \int {d^{} {K^0}
\over \left(2 \pi \right)^{}} {d^{} {{K}} \over
\left(2 \pi \right)^{}} {1\over \left(P_1 - K \right)^0 -  K^2/2
 + i \epsilon}{1\over \left(P_2 + K \right)^0 - K^2/2 + i \epsilon}\nonumber \\
&=&i \left({h^{(0)}\over v}\right)^2 \int {d^{} {K} \over \left(2 \pi
\right)^{}} {1 \over K^2
+\varepsilon}\nonumber \\
&=& i\left({h^{(0)}\over v}\right)^2{\varepsilon^{-1/2}\over 2}.
\end{eqnarray}
Using the matching condition Eq.~(\ref{3.18}), we see that this reproduces the
leading term in Eq.~(\ref{philoopgraph}). It must do this, because one can not
write down an interaction proportional to an inverse fractional power of
$\epsilon$ in the low energy  ${{\rm RNR}\delta}$  effective Lagrangian.   The
${\cal O}(1/v)$ term  in Eq.~(\ref{philoopgraph}) is reproduced in the
effective theory by the  one-loop matching condition to $h$,
\begin{equation}
h^{(1)}=-{3\over 4}{g^4 M\over m^5}=-{3\over 4}{\kappa\over m} h^{(0)}.
\end{equation}
At two loops, one needs to compute the matching conditions in
Fig.~\ref{fig:twoloop}.
The two-loop graph in the effective theory with two insertions of $h^{(0)}$
reproduces the $1/\varepsilon$ term in Eq.~(\ref{phi2loop}), and the one-loop
graph with one insertion of $h^{(0)}$ and one insertion of $h^{(1)}$ reproduces
the $1/\sqrt\varepsilon$ term. The matching correction to $h$ at two loops
requires knowing the coefficients of operators of the form
$\left(\Psi^\dagger \nabla \Psi \right)^2$. This is a complication of the $1+1$
dimensional analysis which is not present in $3+1$ dimensions.

\begin{figure}
\epsfxsize=12cm
\hfil\epsfbox{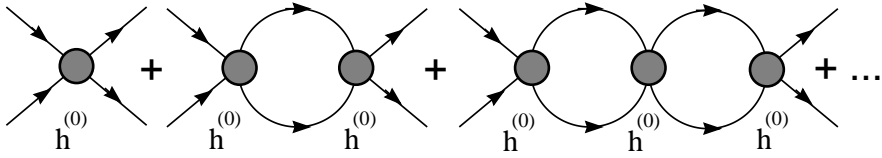}\hfill
\caption{The sum of ladder graphs in ${{\rm NR}\delta}$ using the vertex
$h^{(0)}$ obtained using tree-level matching. \label{fig:dladder}}
\end{figure}

The tree-level matching condition for $h$ only dominates when $\kappa/m\ll 1$.
In this limit, the sum of bubble
graphs in ${{\rm NR}\delta}$ shown in Fig.~\ref{fig:dladder} again gives a
geometric series,
\begin{equation}\label{3.20}
i{h^{(0)}\over v}\ {1 \over 1 - h^{(0)} \varepsilon^{-1/2}/\left(2  v\right) }
\end{equation}
which reproduces the result Eq.~(\ref{philoopgraph}) of ${{\rm RNRY}}$ for
$\kappa\ll m$.

When $\kappa/m$ is not small, we have already seen in the full theory  that the
scattering amplitude is not given by the simple geometric series
(\ref{yladdersum}); the same is of course
true in the effective theory.  Consider a four-fermi operator in the effective
theory containing one time derivative,
\begin{equation}\label{highdim}
{c\over m}\psi^\dagger{\partial_0}\psi \psi^\dagger\psi +\mbox{h.c.}
\end{equation}
Like $h$, higher loop matching conditions to $c$ will be suppressed by powers
of $\kappa/m$, so the matching will have to be done to all orders when
$\kappa/m$ is not small. In the rescaled theory, this operator becomes
\begin{equation}\label{highdimre}
{c\over v}{M v^2\over m}
\Psi^\dagger{\partial_0}\Psi \Psi^\dagger\Psi+\mbox{h.c.}
\end{equation}
In the Born approximation the matrix element of this operator may be  neglected
compared to the terms in the series (\ref{yladdersum}); however, when it is
inserted into a two-loop bubble diagram, it gives a contribution to the
scattering amplitude proportional to
\begin{equation}
{h^2\over v^2}{c\over v}{M v^2\over m}\sim
\left({\kappa\over m}\right)^2{c\over v}
\end{equation}
which, for $\kappa/m\sim 1$, is the same size as the first term in the
geometric series.
Similarly, operators of arbitrarily high order will contribute at all orders
in  $1/v$ to the scattering, suppressed only by powers of $\kappa/m$, so the
effective field theory is unable to describe bound states in the theory when
this parameter is not small.
An
infinite number of operators is required in the effective theory to reproduce
the scattering amplitude in the full theory.
A simple way to see that the
effective theory breaks down is to note that in ${{\rm RNRY}}$, the effective
mass of the $\Phi$ is $m/\sqrt{2}Mv\sim m^3/g^2M=m/\kappa$, for $v\sim
g^2/m^2$. Thus, for $\kappa\ll m$ the $\Phi$ field is heavy, and may be
integrated out, with higher dimension operators suppressed by powers of
$Mv/m\sim\kappa/m$. For $\kappa\sim m$, the $\Phi$ is light, and cannot be
integrated out of the theory if  a light bound states is to be properly
described.

\subsection{$3+1$ Dimensions}

The analysis of the previous subsection can be repeated in $3+1$ dimensions. We
will use the same symbols as in $1+1$ dimensions, but they now have dimension
\begin{equation}\label{3.112}
\left[ \psi \right ] =3/2,\ \left[ \phi \right] =1\ \left[ M \right] = 1,\
\left[ m \right] = 1,\ \left[ g \right] = 0.
\nonumber
\end{equation}
The three dimensional Yukawa potential is
\begin{equation}\label{3.21}
V\left({\bf x} \right) = -g^2\int {d^{3} {{\bf q}} \over \left(2 \pi
\right)^{3}} {e^{i {\bf q} \cdot {\bf x}} \over \left|{\bf q}\right|^2 + m^2} =
-{g^2\over 4 \pi} {e^{-m \left|{\bf x}\right|}\over \left|{ \bf x}\right|}.
\end{equation}
The Schrodinger equation with a three dimensional Yukawa potential does not
necessarily have a bound state. The bound state first appears when
\begin{equation}\label{3.22}
{g^2 \over 4 \pi} {M\over m} \ge 1.7\ \ \ .
\end{equation}
Under the rescaling (\ref{again3.24}), the ${{\rm NRY}}$ Lagrangian in
$3+1$ dimensions becomes
\begin{equation}\label{3.23}
{{\cal L}} = \Psi^\dagger\left( i \partial^0 +  {{\nabla}^2 \over 2 } \right)
\Psi - {g\over\sqrt{v}} \Phi \Psi^\dagger \Psi   + {v\over 2} \left(\partial^0
\Phi \right)^2 -  {1\over 2} \left({\nabla} \Phi \right)^2 - {1\over 2}
{m^2\over M^2 v^2} \Phi^2\nonumber
\end{equation}
while the rescaled version of ${{\rm NR}\delta}$ gives the ${{\rm RNR}\delta}$
Lagrangian
\begin{equation}\label{3.27}
{{\cal L}} = \Psi^\dagger\left( i \partial^0 + { \nabla^2 \over 2 }\right) \Psi
+ h {M ^2 v} \left(\Psi^\dagger \Psi \right)^2.
\end{equation}
The modified scaling rules of Sec.~\ref{sub:modscal} give
\begin{equation}
\left[ x \right ] =- 1,\ \left[ t \right ] = -2,\  \left[ {{\cal L}} \right ]
= 5,\ \left[ \psi \right ] =3/2,\ \left[ \phi \right] =3/2, \label{3.23.1}
\end{equation}
so that $\left[\phi \psi^\dagger \psi \right] - \left[ {{\cal L}} \right] =
-1/2$, and $\left[\left(\psi^\dagger \psi \right)^2 \right] - \left[ {{\cal L}}
\right]=1$, which gives the $v$ dependence of the coefficients in
Eqs.~(\ref{3.23},\ref{3.27}). In three dimensions, the Yukawa interaction is a
relevant operator, and the four-Fermi interaction is an irrelevant operator.
This is a sign that the four-Fermi theory will have trouble correctly
reproducing the behavior of the Yukawa theory when there is a light bound
state in the spectrum.

Tree-level $\Phi$ exchange in ${{\rm RNRY}}$ gives the amplitude for $\Psi\Psi$
scattering
\begin{equation}
I_0=i{g^2 M^2 v\over m^2}+{{\cal O}}(v^3)
\end{equation}
and so the tree level matching condition is the same as in $1+1$ dimensions,
\begin{equation}\label{3.28}
h^{(0)} = {g^2 \over m^2}.
\end{equation}
The one-loop box graph in ${{\rm RNRY}}$  is
\begin{eqnarray}
I_1^Y&=&-i{g^4\over v^2} \int  {d^{3} {{\bf K}} \over \left(2 \pi \right)^{3}}
{1\over  \left|{\bf K}\right|^2 + \mu^2}{1\over\left| {\bf K}+ {\bf P_1}-{\bf
P_1^\prime}\right|^2  + \mu^2 }\times\nonumber\\
&&\qquad {1 \over \left(P_1 + P_2 \right)^0 - \left|{\bf P_2}-{\bf
K}\right|^2/2-  \left|{\bf P_1}+{\bf K}\right|^2/2}\nonumber\\
\noalign{\medskip}
&=& {i\over 8 \pi} {g^4\over v^2}\ {1\over \mu \left( \mu + \sqrt \varepsilon
\right)^2 } \label{3.25}\\ \noalign{\medskip}
& = &i {g^4 v M^3 \over 8 \pi m^3}\left[1-{ 2\varepsilon^{1/2} Mv\over m } +
{{\cal O}}(v^2) \right],\nonumber
\end{eqnarray}
for off-shell incident particles with ${\bf P_1}={\bf P_2}=0$, where
as before $\varepsilon = - \left(P_1+P_2 \right)^0$. The one-loop bubble graph
in the
low-energy theory is
\begin{eqnarray}
I_1^\delta&=&h^2 M^4 v^2 \int {d^{} {K^0} \over \left(2 \pi \right)^{}} {d^{3}
{{\bf K}} \over \left(2 \pi \right)^{3}}  {1\over \left(P_1 - K \right)^0 -
\left|{\bf P_1 - K} \right|^2/2 + i \epsilon} \nonumber \\
&&\times {1\over \left(P_2 + K \right)^0 -  \left|{\bf P_2 + K} \right|^2/2 + i
\epsilon} \\ \noalign{\medskip}
&=&i {h^2 M^4 v^2} \int  {d^{3} {{\bf K}} \over \left(2 \pi \right)^{3}}{1
\over \left|{\bf K}\right|^2 + \varepsilon}\label{3.25.1} \\
\noalign{\medskip}
&=& -i {h^2 M^4 v^2}\ {\varepsilon^{1/2}\over 4 \pi},\nonumber
\end{eqnarray}
where we have evaluated the integral in dimensional regularization. Note that
an integral which diverges as $K^r$, with $r$ an odd integer, is finite when
regulated with dimensional regularization, and needs no subtractions. The
result Eq.~(\ref{3.25.1}) correctly reproduces the leading non-analytic term in
Eq.~(\ref{3.25}). However, the analytic term is absent, so one needs to add a
one-loop matching contribution to $h$, \begin{equation}\label{3.31} h^{(1)} =
{g^4 M  \over 8\pi m^3}={g^2 M\over 4\pi m}\, {1\over 2}\ h^{(0)}.
\end{equation} In the region of parameter space for which bound states exist in
the Yukawa theory, this term is at least as large as the tree-level matching
term $h^{(0)}$, and so all orders in the loop expansion must be calculated in
order to calculate the matching for $h$.  The complete matching condition for
$h$ will be  proportional to the scattering length for a Yukawa potential,
which is not  analytically known in closed form.  Furthermore, there is no
reason for $h$ to be of the naive size $1/m^2$.
In particular, if there is a bound state near threshold, $h$ will be much
larger.

The two-loop ladder graph in the full theory is
\begin{eqnarray}\label{fulltwoloop}
I_2&=&i{g^6\over (4\pi)^2 v^3}\,{1\over\left(\mu^2-\varepsilon\right)^2}\
\ln{(2\mu+\varepsilon^{1/2})^2\over 3\mu(\mu+2\varepsilon^{1/2})}\nonumber \\
\noalign{\medskip}
&=&i{g^6 M^4 v\over(4\pi)^2 m^4}\left[\ln{4\over 3}-{\varepsilon^{1/2} Mv\over
m}+ {\varepsilon M^2 v^2\over 4 m^2}\left(7+8 \ln{4\over 3}\right)+{{\cal
O}}(v^3)\right],
\end{eqnarray}
the two-loop bubble graph in ${{\rm RNR}\delta}$ is
\begin{equation}
i{g^6 M^6 v^3\over (4\pi)^2 m^6} \varepsilon
\end{equation}
and the one-loop graph with a single insertion of $h^{(1)}$ is
\begin{equation}
-2 i h^{(0)}h^{(1)}\, {M^4 v^2\over 4\pi}\, \varepsilon^{1/2} =-i{g^6 M^5
v^2\over(4\pi)^2 m^4}\, {\varepsilon^{1/2}\over m}.
\end{equation}
The sum of the graphs on the right hand side of Fig.~\ref{fig:twoloop} is
\begin{eqnarray}\label{efftwoloop}
i{g^6 M^4 v\over(4\pi)^2 m^4}\left[-{\varepsilon^{1/2} Mv\over m}+ {\varepsilon
M^2 v^2\over 4 m^2}\, 4 \right].
\end{eqnarray}
Comparing Eq.~(\ref{efftwoloop}) with  Eq.~(\ref{fulltwoloop}),  we see that
the nonanalytic term of order $\varepsilon^{1/2}$ is again reproduced in the
effective theory, as it must be, but that the terms of order $\varepsilon$ and
$\varepsilon^0$ terms are not reproduced.  Therefore there is a two loop
contribution to $h$,
\begin{equation}
h^{(2)}={g^6 M^2\over (4\pi)^2 m^4} \ln{4\over 3}=
h^{(0)}\,\left({g^2 M\over 4\pi m}\right)^2 \ln{4\over 3},
\end{equation}
which once again is at least as large as $h^{(0)}$ in the region of interest,
while at order $\varepsilon$ the difference between the graphs in the two
theories
\begin{equation}
i{g^6 M^4 v\over(4\pi)^2 m^4}\ {\varepsilon M^2 v^2\over 4 m^2}\left(3+8
\ln{4\over 3}\right)
\end{equation}
contributes to the matching conditions of an operator such as $\Psi^\dagger
i\partial^0 \Psi \Psi^\dagger \Psi$, with coefficient
\begin{equation}
{g^6 M^6 v^3\over (4\pi)^2 m^6}\, \left({3\over 4}+ 2\ln{4\over 3}\right)=
\left({3\over 4}+ 2\ln{4\over 3}\right)\, \left({h^{(0)}M^2 v \over 4\pi}
\right)^2\, h^{(0)} M^2 v.
\end{equation}
The important observation is that this counterterm contributes to the
$\varepsilon$ term at order $(QM)^3$, the same order as
the two-loop bubble graph in the
effective theory; hence, without knowing the coefficients of the dimension
eight operators in the effective theory all terms of the form
$(QM)^n$ are not summed.   It is clear that a similar situation exists at
higher  loops:
each loop graph in the effective theory is the same order as a counterterm, and
so without knowing the counterterms, the graphs may not be summed.  Thus,
without
including all higher dimension operators, the effective theory does not
correctly
sum all terms of order $(QM)^n$.

\subsection{Bound States and The Effective Range Expansion}\label{sub:ere}

Let us look for bound state poles in the scattering amplitude. In 1+1
dimension, the bubble chain sum in the effective theory had contributions
proportional to inverse powers of $v$, which diverged in the low energy limit.
Thus, for weak coupling there was a bound state at threshold, which was
reproduced in the low-energy theory. In 3+1 dimensions, the full theory with a
Yukawa potential has a bound state at threshold when $g^2 M/4\pi m\simeq 1.7$.
Since in the bubble sum in ${{\rm NR}\delta}$ only positive powers of $v$
occur, all higher order graphs contributing to the scattering amplitude vanish
at low energy.  Therefore, there can be no bound state pole at threshold unless
one of the coefficients in the  effective Lagrangian ${{\rm NR}\delta}$
diverges. In particular, there can be a bound state at threshold if $h
\rightarrow \infty$.    To see this in the $\delta$-function
effective theory, one must evaluate the matching condition for $h$ to all
orders, since $g^2 M/4\pi m$ is not small.

For finite $h$ the bubble sum in the effective theory may be summed to find a
bound
state for finite $h$. Neglecting the contributions from
operators of dimension $>6$, the bubble chain sum is
\begin{eqnarray}\label{3.30.1}
&&i {h M^2 v} \left[ 1 - {h M^2 v\varepsilon^{1/2} \over 4 \pi } + \left({h M^2
v\varepsilon^{1/2} \over 4 \pi } \right)^2 + \ldots \right]\nonumber \\
\noalign{\medskip}
&&=i {h M^2 v}\ {1 \over 1 + h M^2 v\varepsilon^{1/2} /4 \pi }.
\end{eqnarray}
which is the expression given by Weinberg~\cite{weinberg}. There is a pole at
\begin{equation}\label{3.31.1}
{h M^2 v\varepsilon^{1/2} \over 4 \pi}  = -1,
\end{equation}
or rescaling back to physical units, at
\begin{equation}\label{3.32}
{h M^{3/2} \left(-E \right)^{1/2} \over 4 \pi}  = -1.
\end{equation}
which, for $h<0$, has a bound state below threshold.  The bound state energy is
\cite{weinberg}
\begin{equation}\label{3.33}
E = - {16 \pi^2 \over h^2 M^3}.
\end{equation}
When $|h|\rightarrow\infty$, the
bound state approaches threshold.

We have argued, however, that the bubble sum in the effective theory does not
correctly
sum the terms of order $(QM)^n$.   What relation does this bound state then
have to a bound state in the full theory?  To answer this, it is useful to
recall some results of the effective
range expansion for potential scattering.

The scattering
amplitude $i{\cal A}$ is related to the  phase-shift $\delta$ by
\begin{equation}\label{3.36}
k \cot \delta = {4 \pi \over M {\cal A}} + i k
\end{equation}
In the effective range expansion, $k\,\cot \delta$
is expanded in powers of $k$,
\begin{equation}\label{3.38}
k \cot \delta = -{1\over a} + {1\over 2 } r_e k^2 +
c_4 k^4 + \ldots.
\end{equation}
This expansion gives us some useful additional information, because it is known
to have
a radius of convergence $\ge m/2$ \cite{gw,taylor}. Here $a$ is the
scattering length, and $r_e$ is the effective range. The first
two terms in this series provide a very good approximation to the measured
nucleon scattering cross-sections. As shown in
Ref.~\cite{ksw},
Eqs.~(\ref{3.36}) and (\ref{3.38}) imply that the four-Fermi operators in
${\rm NR}\delta$ have a momentum expansion of the form
\begin{equation}\label{3.100}
-{4 \pi \over M} \left[  a + {1\over 2} a^2 r_e k^2 +
\left(a^3 {1\over 4} r_e^2 - a^2 c_4\right) k^4
+\ldots \right].
\end{equation}
Rescaling these coefficients, we find that $h$ is related to the
exact  scattering length $a$ of the Yukawa theory,
\begin{equation}\label{3.30.2}
h  = - {4 \pi \over M} a
\end{equation}
whereas a two-derivative term like $\Psi^\dagger{\bf \nabla}^2\Psi
\Psi^\dagger\Psi$ has a coefficient of order
\begin{equation}\label{twoder}
2\pi M^3 v^3 a^2 r_e.
\end{equation}
The scattering length diverges when there is a bound state at threshold, so
that $h \rightarrow \infty$.  It is known from potential scattering that the
other
coefficients in the effective range expansion (\ref{3.38}) do not
diverge~\cite{gw,taylor}.
In this case, the coefficients of the higher dimension operators diverge as
well, as is evident in Eq.~(\ref{3.100}), and the divergent behavior of the
various coefficients is highly correlated.   It is difficult to see this
behaviour by
studying the matching conditions to the effective theory, since one must
perform
the matching to all orders in the loop expansion.  However, the effective range
expansion gives nontrivial information on the form of the complete matching
conditions.

If we now consider $\psi\psi$ scattering, the two-bubble graph from the
dimension six operator $\left(\psi^\dagger\psi\right)^2$ gives a term of order
\begin{equation}
{a^3 M^3 v^3\over 4\pi}
\end{equation}
which is to be compared with the counterterm contribution (\ref{twoder}).
The terms have the same explicit dependence on $M$ and $v$, as we have already
argued,
so the bubble graph is the same order as the counterterm using the power
counting
scheme of \cite{weinberg,vka,vkb,ksw}.
Furthermore both terms are suppressed
by two powers of $Q=Mv$ relative to the contact term, so even when $M$ is large
it
is not necessary to resum the bubble graphs in this scheme.  However, the
bubble graph will
dominate when
\begin{equation}
{a\over 4\pi r_e}\gg 1.
\end{equation}
Summing bubble graphs in the effective range expansion therefore corresponds
to an expansion in powers of
$a/4\pi r_e$, not $Q/M$.   Therefore, when the scattering length is large (and
therefore
the bound state is nearly at threshold) the sum of the bubble graphs in
NR{$\delta$}
correctly describes the bound state, and (\ref{3.33}) is valid, up to
corrections of
order $4\pi r_e/a$.
However, as discussed in \cite{ksw}, when the scattering length is large the
effective theory breaks down not at $k\sim m\sim r_e^{-1}$, but at a much lower
scale
\begin{equation}
k\sim \sqrt{2 \over a r_e}
\end{equation}
which, in the region where the bubble graphs dominate, is much less than $m$.
The region of convergence is not controlled by the scalar mass, and vanishes
as the bound state approaches threshold.    Thus, we still do not have the
situation
we had in weakly bound 1+1 dimensional theory in which the bound
state was correctly predicted in the low-energy theory but the radius of
convergence
of the theory was set by the heavy scalar mass.

\subsection{Implications for Chiral Perturbation Theory}
The results are unchanged when pions are included in the chiral Lagrangian,
since pion exchange scales with $M$ and $v$ in the same way as the dimension
six  contact  interactions.
We may therefore conclude from this simple model that chiral perturbation
theory
cannot sum all terms of order $(QM)^n$ without including an infinite number of
higher
dimension operators.    Using the effective range expansion, we see that the
bubble graphs
only correctly describe the low-energy bound state when the effective field
theory breaks
down at a scale much less than the symmetry-breaking scale $\Lambda_{\chi SB}$
(the
analogue of the scalar mass in our simple model).  Therefore
the standard form of heavy nucleon chiral perturbation theory does not provide
the correct
description of nucleon-nucleon scattering up to energies set by the scale of
the heavy
excitations which have been integrated out of the theory.   This does not mean
that the effective field theory idea is useless:
it does mean, however, that the usual chiral Lagrangian is not the correct
effective
field theory.  If the appropriate low energy degrees of freedom are introduced
by hand as new fields
in the effective Lagrangian, it should be possible to correctly describe low
energy
nucleon-nucleon scattering in an effective field theory.\footnote{We thank D.
Kaplan for
discussions on this point.}  However, the properties of the low energy
degrees of freedom are not determined by the parameters in the standard
chiral Lagrangian, unless all higher dimension operators are included.

\section{Conclusions}\label{sec:conc}

Rescaled non-relativistic effective field theories simplify the study of weakly
bound states in quantum field theory. The use of RNRQCD makes the $v$ power
counting of NRQCD manifest, and reorganizes the perturbation expansion in a
more systematic way. Rescaled effective theories can also be used to study
Yukawa bound states due to scalar exchange.  In $1+1$ dimensions, a bound state
occurs at weak coupling because the $n$-loop ladder graph diverge is
proportional to $1/(-E)^{n/2}$. In this limit one may sum the diagrams of
Fig.~\ref{fig:dladder} and obtain the properties of the bound state, without
worrying about higher loop matching conditions. In three dimensions, one
needs a critical coupling before there exists a bound state, and the problem is
intrinsically strongly coupled. One needs to include the full Yukawa
interaction to study the bound state. Equivalently, the matching from ${{\rm
NRY}}$ to ${{\rm NR}\delta}$ must be performed to all orders in the loop
expansion, and the higher order matching terms are relevant for the bound
state. We have shown by explicit computation that the power counting scheme of
Refs.~\cite{weinberg,vka,vkb,ksw} that involves summing powers of $QM$
does not hold for Yukawa theory.
One can not, in general replace the Yukawa potential by a
$\delta$-function potential to study even weakly coupled bound states in three
dimensions, except for a very limited region in momentum space $\left|{\bf
k}\right| < \sqrt{2/a r_e}$, which vanishes as the bound state approaches
threshold.

\acknowledgments
We would like to thank D.B.~Kaplan for many useful discussions concerning
this work.  We also thank C.~Bauer,   M.J.~Savage, B. Smith and
M.B.~Wise for useful discussions. This work was supported in part by a
Department of
Energy grant DOE-FG03-90ER40546, by a Presidential Young Investigator award
PHY-8958081 from the National Science Foundation, by the Alfred P. Sloan
Foundation and by the Natural Sciences and Research Council of Canada.
Some of this work was done at  the Benasque Centre for Physics, and we thank
the
organizers of the centre for their hospitality.

\end{document}